\documentclass[11pt,preprint]{aastex}

\usepackage{amssymb}
\usepackage{amsmath}

\newcommand{\SgrA}{Sgr A$^{\star}\,$}

\shorttitle{{\it NuSTAR} detection of Sagittarius A$^{\star}$ hard X-ray flares}
\shortauthors{Barri{\`e}re et al.}

\begin{document}


\title{NuSTAR detection of high-energy X-ray emission and rapid variability from Sagittarius A$^{\star}$ flares}


\author{
Nicolas~M. Barri{\`e}re\altaffilmark{1},
John~A. Tomsick\altaffilmark{1}, 
Frederick~K. Baganoff\altaffilmark{2}, 
Steven~E. Boggs\altaffilmark{1},
Finn~E. Christensen\altaffilmark{3},
William~W. Craig\altaffilmark{1,4}, 
Jason~Dexter\altaffilmark{5},
Brian Grefenstette\altaffilmark{6},
Charles~J. Hailey\altaffilmark{7},
Fiona~A. Harrison\altaffilmark{6}, 
Kristin~K. Madsen\altaffilmark{6},  
Kaya Mori\altaffilmark{7},
Daniel Stern\altaffilmark{8},
William~W. Zhang\altaffilmark{9}, 
Shuo~Zhang\altaffilmark{7},
Andreas Zoglauer\altaffilmark{1} 
}

\altaffiltext{1}{Space Sciences Laboratory, University of California, Berkeley, CA 94720, USA.}
\altaffiltext{2}{Kavli Institute for Astrophysics and Space Research, Massachusetts Institute of Technology, Cambridge, MA 02139-4307, USA.}
\altaffiltext{3}{National Space Institute, Technical University of Denmark, Copenhagen, Denmark.}
\altaffiltext{4}{Lawrence Livermore National Laboratory, Livermore, CA 94550, USA.}
\altaffiltext{5}{Departments of Physics and Astronomy, University of California, Berkeley, CA 94720, USA.}
\altaffiltext{6}{Cahill Center for Astronomy and Astrophysics, Caltech, Pasadena, CA 91125, USA.}
\altaffiltext{7}{Columbia Astrophysics Laboratory, Columbia University, New York, NY 10027, USA.}
\altaffiltext{8}{Jet Propulsion Laboratory, California Institute of Technology, Pasadena, CA 91109, USA.}
\altaffiltext{9}{X-ray Astrophysics Laboratory, NASA Goddard Space Flight Center, Greenbelt, MD 20771, USA.}


\begin{abstract}
Sagittarius A$^{\star}$ harbors the supermassive black hole that lies at the dynamical center of our Galaxy.  Sagittarius A$^{\star}$ spends most of its time in a low luminosity emission state but flares frequently in the infrared and X-ray, increasing up to a few hundred fold in brightness for up to a few hours at a time. The physical processes giving rise to the X-ray flares are uncertain. Here we report the detection with the {\em NuSTAR} observatory in Summer and Fall 2012 of four low to medium amplitude X-ray flares to energies up to 79 keV. For the first time, we clearly see that the power-law spectrum of Sagittarius A$^{\star}$ X-ray flares extends to high energy, with no evidence for a cut off. Although the photon index of the absorbed power-law fits are in agreement with past observations, we find a difference between the photon index of two of the flares (significant at the 95\% confidence level). The spectra of the two brightest flares ($\sim$55 times quiescence in the 2--10 keV band) are compared to simple physical models in an attempt to identify the main X-ray emission mechanism, but the data do not allow us to significantly discriminate between them.  However, we confirm the previous finding that the parameters obtained with synchrotron models are, for the X-ray emission, physically more reasonable than those obtained with inverse-Compton models. One flare exhibits large and rapid ($<$ 100~s) variability, which, considering the total energy radiated, constrains the location of the flaring region to be within $\sim$10 Schwarzschild radii of the black hole.
\end{abstract}





\section{Introduction}
\label{sec:intro}
Sagittarius A$^{\star}\,$ (\SgrA), at a distance of $\sim8$~kpc, is a $\sim4 \times10^6$ solar mass supermassive black hole (SMBH) that marks the dynamical center of the Milky Way \citep{ghez.2008ys,gillessen.2009vn}.  Although \SgrA is a fairly bright radio and submillimeter (sub-mm) source, it is not detected in the optical and UV due to over 30 magnitudes of visual extinction, and it is very dim in X-rays. Its bolometric luminosity $L_{\rm bol}$ is only 550 times that of the Sun, corresponding to $\sim10^{-9}$ times the Eddington luminosity \citep{narayan.1998kx}.  Most of this is radiated in the sub-mm, with the X-ray luminosity being $3.6 \times 10^{33}$ erg s$^{-1}$ \citep{nowak.2012fk} in the 2--10 keV energy band ($\sim 2 \times 10^{-3} \, L_{\rm bol}$). This low luminosity implies that matter falls onto the black hole in a radiatively-inefficient manner, perhaps in a hot, geometrically thick, optically thin flow \citep{narayan.1998kx,yuan.2003qf,wang.2013fk}. 

A few times per day in the near infrared (NIR) and about once per day in X-rays, \SgrA flares up to a few hundred times its quiescent level for intervals lasting up to a few hours, with the strong flares being less common than the weak ones. The NIR emission is strongly polarized \citep{eckart.2006kl}, indicating a synchrotron origin, and therefore a population of relativistic electrons. The source of the X-ray emission is uncertain; possibilities include both synchrotron radiation with a cooling break \citep[SB,][]{yuan.2003qf, yuan.2004uq, dodds-eden.2009dq} and inverse Compton (IC) up-scattering of lower energy photons by energetic electrons. Two IC scenarios involve the NIR emitting electrons up-scattering either ambient sub-mm photons \citep[external Compton, EC,][]{markoff.2001ve,eckart.2004oq,yusef-zadeh.2006ij} or the NIR synchrotron emission itself \citep[synchrotron self-Compton, SSC, e.g.][]{markoff.2001ve,eckart.2006zr,marrone.2008hc}. A third IC scenario involves NIR flare photons up-scattered by the electrons radiating in the sub-mm domain \citep{yusef-zadeh.2009fk,yusef-zadeh.2012fk}.

The physical cause of the episodic particle acceleration is not understood, and suggestions include a hot spot in the accretion flow \citep[due, for example, to a magnetic reconnection event, see e.g.][]{dodds-eden.2010fk}, enhanced mass accretion \citep{tagger.2006uq}, magnetohydrodynamic turbulence or shocks in the inner accretion flow \citep{yuan.2003qf}, a jet \citep{markoff.2001ve}, or episodic outflow triggered by magnetic reconnection \citep{yuan.2009fk}. Tidal disruption of asteroids has also been proposed as possible origin of the flares \citep{cadez.2008uq, kostic.2009kx, zubovas.2012fk}.

Numerous flares have now been observed in X-rays \citep{baganoff.2001qy,porquet.2003bh,porquet.2008uq,degenaar.2013cr,neilsen.2013fk}, in NIR \citep{genzel.2003fj,witzel.2012dq}, and in both bands simultaneously \citep{eckart.2004oq,yusef-zadeh.2006ij,eckart.2006zr,marrone.2008hc,dodds-eden.2009dq,trap.2011fk,eckart.2012bh}. Above 10~keV, only upper limits exist due to a combination of limited instrumental sensitivity and spatial resolution \citep{yusef-zadeh.2009fk,trap.2011fk}.
Short, high-amplitude temporal substructures, with variability timescales as short as 47~s, 
have been seen during NIR flares \citep{dodds-eden.2009dq}, indicating a compact origin for this emission component. In the X-ray, variability on 100--200~s timescales has been reported, although at low relative amplitude \citep{porquet.2003bh, nowak.2012fk}. The fastest timescale observed to-date for significant (more than a factor of a few) X-ray variability is 600~s \citep{baganoff.2001qy}.

We report here on {\it NuSTAR} observations of four flares detected in Summer and Fall 2012. This paper is organized as follows. Section \ref{sec:observations} details the observations, and section \ref{sec:BB} presents the method to search for flaring activity. In section \ref{sec:variability}, we quantify the variability amplitude and time scale exhibited during one of the flares. Section \ref{sec:spectralanalysis} describes the spectral analysis of the flares. Then, in section \ref{sec:sed}, we introduce the models that are compared to the SEDs of the two brightest flares, and the results are discussed in section \ref{sec:discussion}. Finally, in section \ref{sec:energybalance}, we derive a constraint on the location of the flaring region.

\section{Observations}
\label{sec:observations}
The {\em NuSTAR} high-energy X-ray observatory \citep{harrison.2013ly} observed the Galactic center three times in Summer and Fall 2012 as part of a coordinated campaign with the {\em Chandra} and Keck observatories: from July 20, 2:11 to July 23, 19:11, from August 04, 07:56 to August 06, 01:06, and from October 16, 18:31 to October 18, 01:26 (UT time). Flaring activity from \SgrA was detected by {\em NuSTAR} on July 20 and 21, and on October 17. Four flares were detected, which are referred to hereafter by J20, J21\_1, J21\_2 and O17.

Unfortunately, only O17 was simultaneously observed by {\em Chandra}, and none of the flares were covered by Keck. We report in this paper on the {\em NuSTAR} data only.  The {\em Chandra} data for the O17 flare will be presented in a future work.

\begin{figure}[h]
\centering
\includegraphics[width=0.6\textwidth]{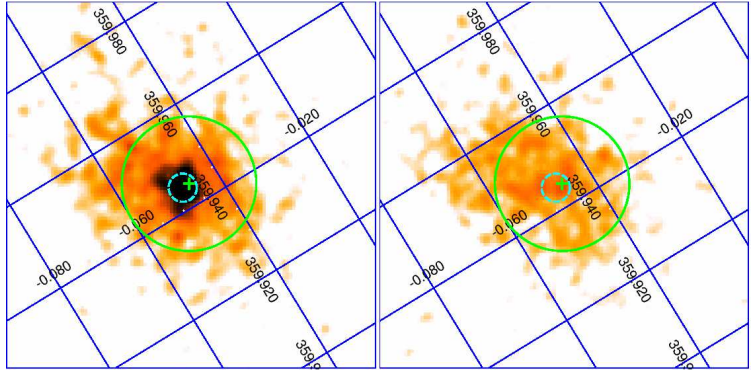}
\caption{The \SgrA field ($4.5' \times 4.5'$) during flare J21\_2 (left panel) and before the flare (right panel) as seen by {\em NuSTAR}'s focal plane module A. The images are made with 3 to 79~keV photons without background subtraction and an exposure time of 3000~s for each panel. It is smoothed with a 3-pixel radius Gaussian function. The image is oriented in equatorial coordinates with North to the top and East to the left, with a logarithmic color scale. The grid shows Galactic coordinates, the cross shows the Reid et al. (2004) radio position of \SgrA, and the small circle (dash line) shows the 3-$\sigma$ {\it NuSTAR} error circle. The large circle shows our $50''$ radius extraction region, centered on the radio position.}
\label{fig:flare_image}
\end{figure}

\section{Flare search: Bayesian block analysis}
\label{sec:BB}

\begin{deluxetable}{cccccc}
\tabletypesize{\small}
\tablecaption{Flares IDs, start and stop times (UT), duration, and significance (for FPM A and B combined).}
\tablewidth{0pt}
\tablehead{
\colhead{Flare ID} & \colhead{UT Date} & \colhead{Start} & \colhead{Stop} & \colhead{Duration (s)} & \colhead{Level ($\sigma$)}  
}
\startdata
J20 & 2012 July 20 & 12:15:21\tablenotemark{a} & 12:30:39  & 920\tablenotemark{a} & 4.8 \\
J21\_1 & 2012 July 21 & 01:45:15\tablenotemark{b} & 02:06:14\tablenotemark{b} & 1238\tablenotemark{b} & 7.5 \\
J21\_2 & 2012 July 21 & 06:01:12\tablenotemark{a} & 06:52:50 & 3099\tablenotemark{a}&  19.7 \\
O17 & 2012 October 17 & 19:50:08\tablenotemark{b} & 20:10:56\tablenotemark{b} & 1249\tablenotemark{b} & 20.2 \\
\enddata
\tablenotetext{a}{This flare was truncated at the beginning by Earth occultation.}
\tablenotetext{b}{This flare was truncated at the beginning by a passage in the SAA and at the end by Earth occultation.}
\label{tab:flareID}
\end{deluxetable}

The data analysis was done with the {\it NuSTAR Data Analysis Software (NuSTARDAS)} v1.1.1.
We first screened the data for South Atlantic Anomaly (SAA) passages and other flaring particle backgrounds. 
Then, we used an extraction region  with a $50''$ radius (encompassing  $\sim$ 68\% of the PSF) centered on the radio position of \SgrA \citep{reid.2004fk} to extract the events from focal plane modules (FPM) A and B in the full energy band (3 -- 79 keV), and we performed a Bayesian block analysis of the combined light curves to search for flares. This analysis is applied on the raw events of both FPMs after the data gaps due to Earth occultation and SAA passages are suppressed. We correct the exposure by applying a coefficient to each event corresponding to the deadtime at the time of the event. The Bayesian block analysis assumes that the light curve can be modeled by a succession of constant rate blocks. The rate within each block is simply calculated based on the number of events it contains and its duration. 
The algorithm to find the optimal location of the change points is described in \citet{scargle.2013vn}. The number of change points is affected by two input parameters: the false positive rate $fpr$ and the prior estimate of the number of change points, $n_{\rm cp\_prior}$. We used $fpr = 0.01$ and a geometric prior: 
$n_{\rm cp\_prior}= 4 - \log( fpr / ( 0.0136 \, N^{0.478} ) )$, where $N$ is the total number of events \citep{scargle.2013vn}.

We investigated the actual false positive rate by performing simulations with synthetic datasets having a constant input rate (equal to the average rate of our dataset) and following a Poisson distribution for the event arrival time intervals. We obtain identical results for both datasets (the July and October light curves have 52,109 and 28,067 events, respectively). We find that spurious peaks (a peak is made of two change points) are detected 0.4\% of the time (1000 trials performed). See Appendix \ref{sec:appendix1} for more details on the sensitivity of the search.

\begin{figure}[p]
\centering
\includegraphics[width=0.9\textwidth]{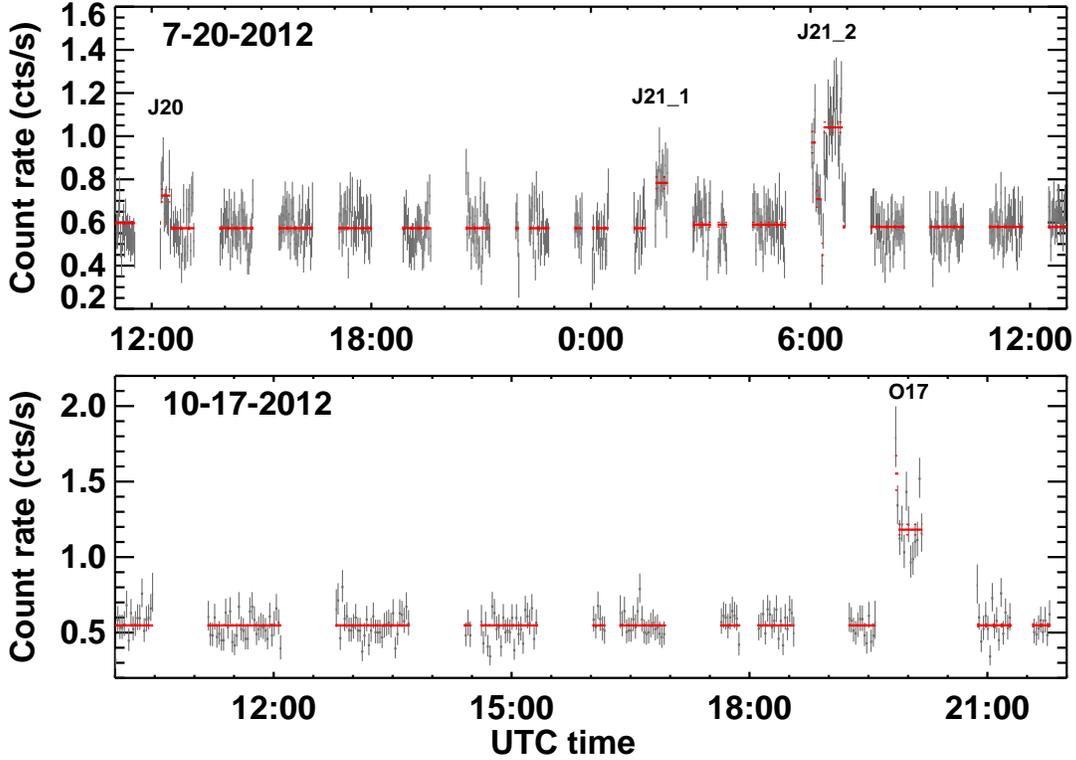}
\caption{Light curves from a $50''$ radius region centered on the radio position of \SgrA in the 3--79~keV energy band for {\em NuSTAR}'s modules A and B combined. The top panel shows a portion of the July observation, and the bottom one shows a portion of the October observation. In both panels, the grey histogram has 100 s time resolution and shows 1-$\sigma$ upper and lower count rate limit. The red lines show the Bayesian block representation of this light curve \citep{scargle.2013vn}, with 1-$\sigma$ rate limits as dotted lines. Instrumental background, estimated to contribute $< \, 5\times10^{-3}$~cts/s in this energy band for this extraction region, has not been subtracted in this plot. The diffuse emission from the \SgrA complex is the main component to the $\sim$ 0.6~cts/s that constitutes the baseline of the light curve. The flares IDs are shown in the plot.}
\label{fig:full_lc}
\end{figure}

\begin{figure}[p]
\centering
\includegraphics[width=0.8\textwidth]{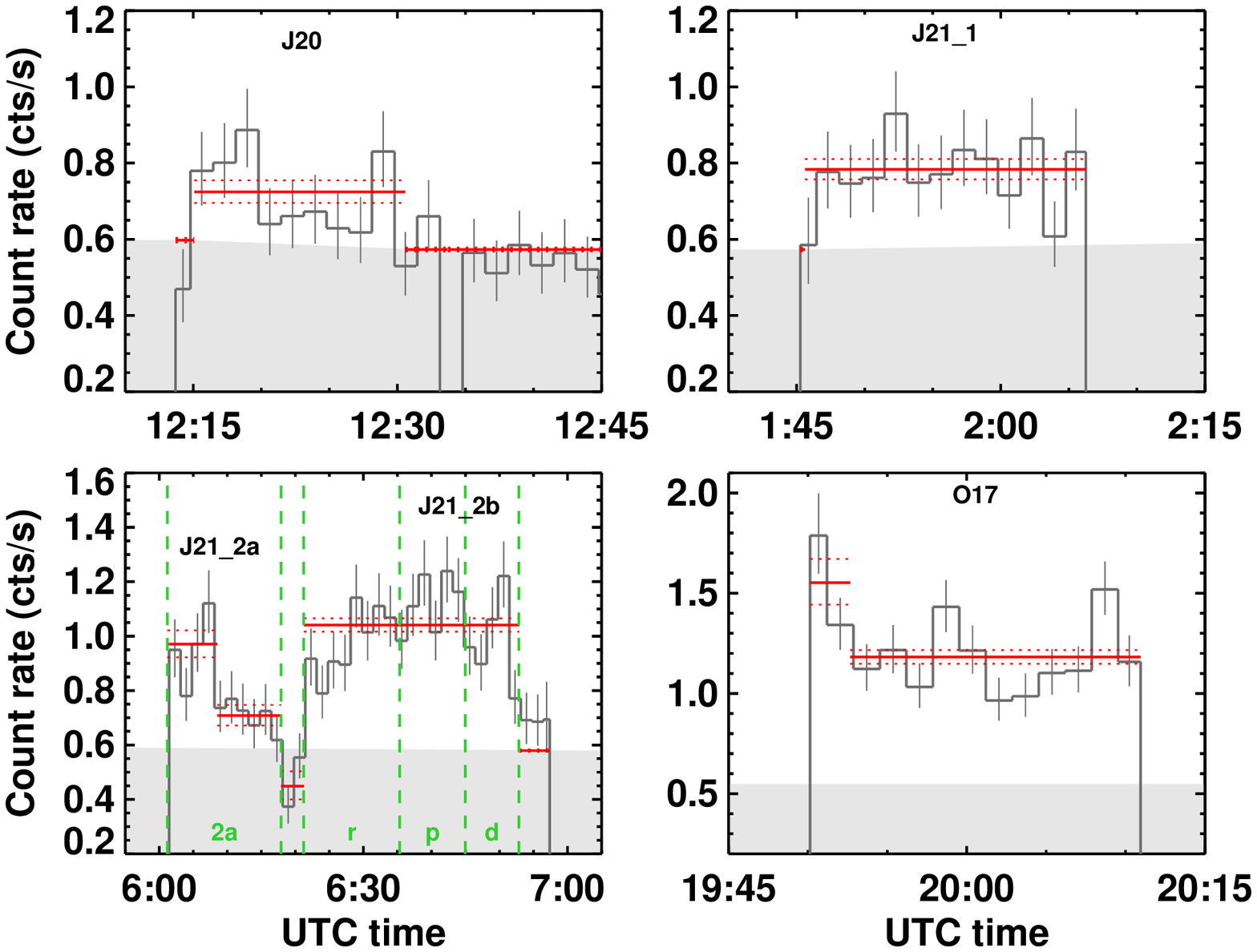}
\caption{Close up view of the light curves and associated Bayesian block representations for the four flares reported in this paper. Binning and error bars are identical to those in Figure \ref{fig:full_lc}. The shaded areas show the baseline count rate as defined by the bayesian blocks preceding and following each flare.}
\label{fig:zoom_lc}
\end{figure}

This analysis led to the detection of four flares as listed in Table \ref{tab:flareID}. The flare blocks are used to define the flaring times. The quiescent time is defined independently for each observation and includes the time remaining after removing the flares and the flaring background events. The quiescent time for the July and October observations amounts to 53.5 ks and 41.8 ks, respectively. Figure \ref{fig:full_lc} shows the July and October portion of the light curve where the flares were detected, and Figure \ref{fig:zoom_lc} shows close-up views of the flare light curves. {\em NuSTAR} is in low-Earth orbit, and the gaps in the data are due to Earth occultation and passages through the SAA.

The position of the flaring source is $\alpha_{\rm J2000} = 17^h45^m40.4^s$,  $\delta_{\rm J2000} = -29^{\circ}00'31.04''$, with an error circle of $10.5''$ radius (3-$\sigma$ confidence level), including a systematic component estimated to be at the $5''$ level (Figure \ref{fig:flare_image}). The position is consistent with the radio location of \SgrA \citep[$5.6''$ offset;][]{reid.2004fk}. The association with \SgrA is clear; no variable source was detected by {\em Chandra} within the error circle during the July 20-23  and October 17 observing campaigns (F. Baganoff, private communication), and the characteristics of the flares are similar to those seen in prior low-energy X-ray observations. No other source is known to produce X-ray flares with luminosity of the order of $10^{35}$ erg/s, a duration of the order of an hour, and a power-law spectrum with photon index around 2.

\section{July flare variability}
\label{sec:variability}
Substructures are clearly visible in the light curves of flares J21\_2 and O17. In particular, on 2012 July 21, there is a rise at 6:21 UT and a decay at 6:51 UT that both happen within one 100-s bin (Figure \ref{fig:zoom_lc}). These features are visible in both FPM A and FPM B (see Appendix \ref{sec:J21_2}). To estimate the amplitude of these variations, it is necessary to understand that the light curve is offset due to the unresolved high-energy emission that is present within our $50''$ radius extraction region, and that this diffuse emission dominates the count rate at all times except during flares J21\_2 and O17. This baseline count rate is measured to be $0.585 \pm 0.011$ cts/s, as averaged using the blocks preceding and following flare J21\_2.

We first focus of the rise at 6:21 UT. Estimating its amplitude is not straightforward as the bottom point is at the level of the quiescent flux (a statistical fluctuation makes it appear below the quiescent level, visible in the bottom left panel of Fig. \ref{fig:zoom_lc}), consistent with a non-detection during the dip that separates sub-flares J21\_2a and J21\_2b. Near the time of the rise, the bottom and top bins show rates of $0.55^{+0.09}_{-0.08}$ cts/s and $0.92^{+0.11}_{-0.10}$  cts/s, respectively. We use the upper 1-$\sigma$ error bar as the upper limit of the flare count rate during the minimum, which leads to a lower limit on the rise factor of $3.8 \pm 1.1$. During the 6:51 UT decay, the count rate drops from $1.22^{+0.13}_{-0.12}$ cts/s to $0.77^{+0.10}_{-0.09}$ cts/s. Subtracting the baseline count rate from both these values, we find a decay factor of $3.4 \pm 1.8$.

\section{Spectral analysis}
\label{sec:spectralanalysis}

\begin{figure}[p]
\centering
\includegraphics[width=0.6\textwidth]{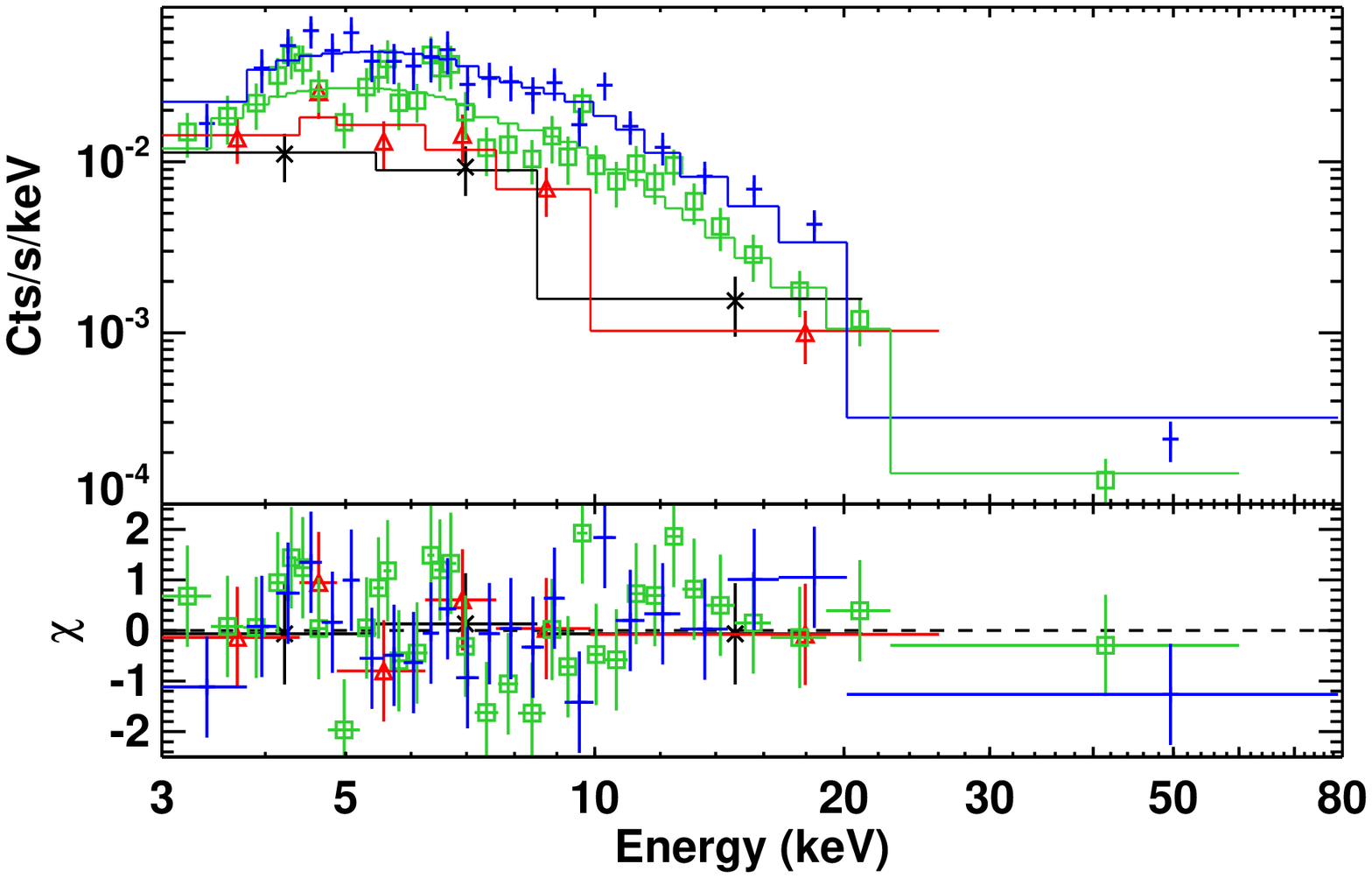}
\caption{ Counts spectrum of the four flares extracted from a $50''$ radius circular region centered on the \SgrA radio position: J20 (black crosses), J21\_1 (red triangles), J21\_2 (green squares), and O17 (blue plus signs). Each spectrum is the combination of data from modules A and B. The data points are shown with 1-$\sigma$ error bars, and the solid lines show the best fit model. The lower panel shows the deviation from the model in units of standard deviation. }
\label{fig:flares_spectrum}
\end{figure}

\begin{figure}[p]
\centering
\includegraphics[width=0.5\textwidth]{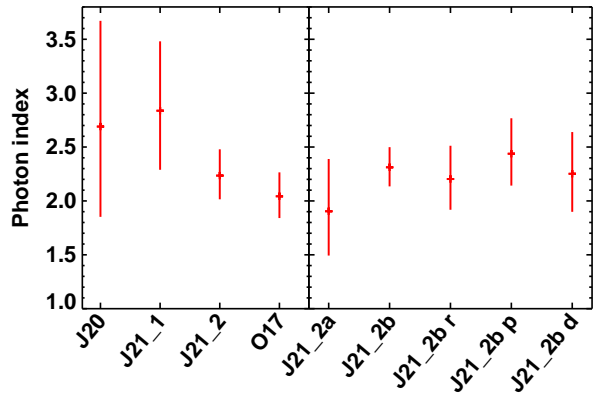}
\caption{Photon index ($\Gamma$) of the flares (left panel), and for a breakdown of flare J21\_2 (right panel), with the letters r, p, and d referring to rise, peak and decay, respectively. In the left panel, the four flares are fit jointly by an absorbed power-law model with the column density tied ($1.66^{+0.70}_{-0.61} \times 10^{23}$ cm$^{-2}$). The breakdown of flare J21\_2 uses a column density fixed to the value determined using the four flares. The errors bars show the 90\% confidence range. }
\label{fig:photindex}
\end{figure}

\begin{figure}[p]
\centering
\includegraphics[width=0.5\textwidth]{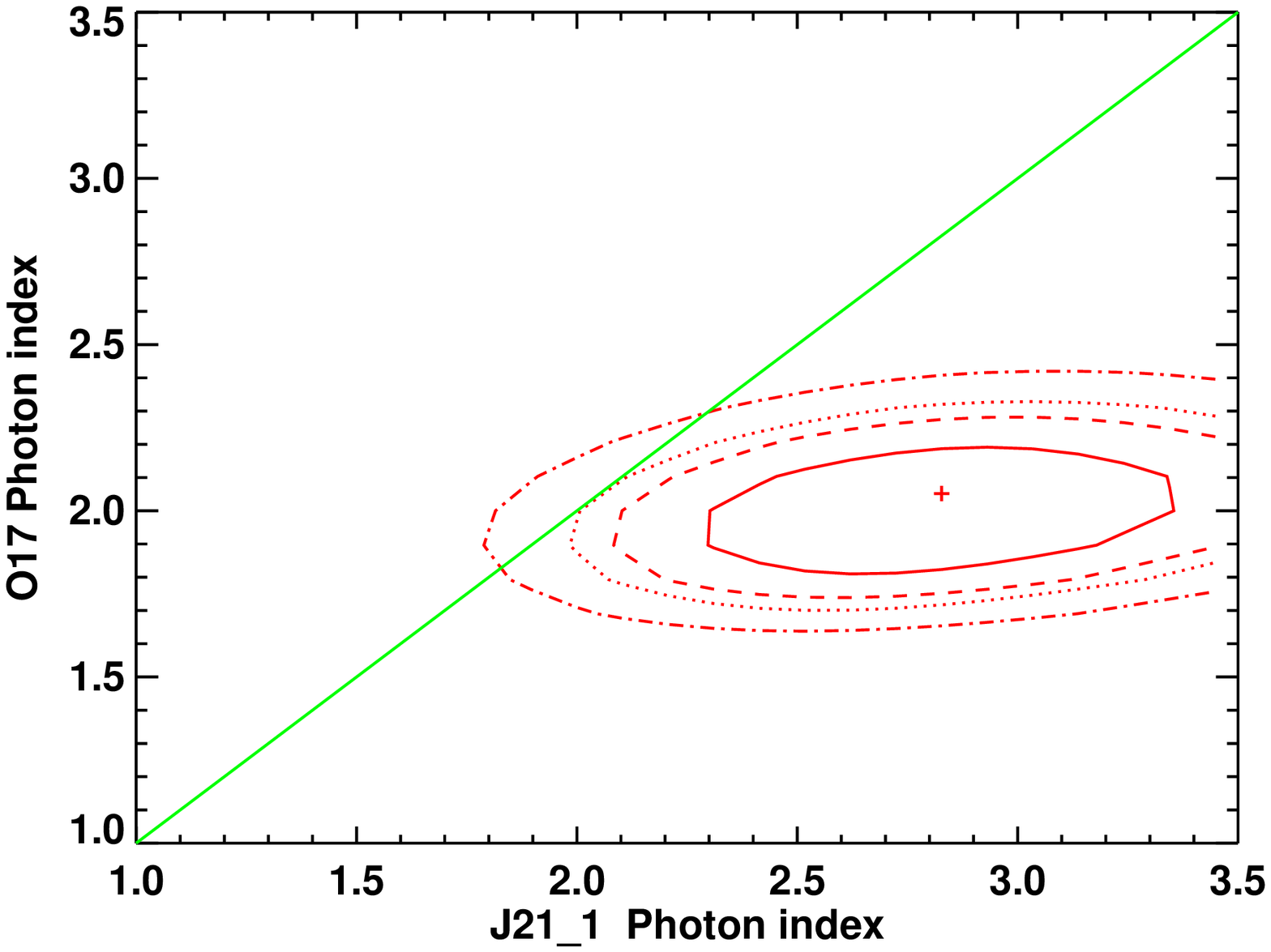}
\caption{Confidence contours for the photon indices of flares J21\_1 and O17, determined while fitting simultaneously the four flares with tied column density. The four contours show the 68.3\%, 90\%, 95\% and 99\% confidence levels (solid line, dash line, dot line, and dash-dot line, respectively). The straight line shows the values where the photon indices of both flares are equal. One can see that the photon index of flares O17 and J21\_1 are different at the 95\% level.}
\label{fig:contourplot}
\end{figure}

The flare and quiescent spectra are extracted from the same $50''$ radius region centered on the radio position of \SgrA. The analysis is done with the Interactive Spectral Interpretation System \cite[ISIS,][]{houck.2000fk} by setting the non-flare spectra as background to the flare spectra.
The model $\texttt{TBabs} \times \texttt{PEGpwrlw}$ (i.e., an absorbed power-law) is used to fit the data, with \citet{verner.1996kx} atomic cross sections  and \citet{wilms.2000vn}  abundances  for the interstellar absorption. 

Some previous studies of  the spectra of \SgrA flares included a dust scattering correction \cite{baganoff.2001qy,porquet.2003bh,porquet.2008uq,nowak.2012fk}. The effect of dust is twofold; it contributes to absorption (through the photoelectric effect), which is accounted for by $\texttt{TBabs}$ in our model, and it scatters X-rays, primarily in the forward direction, creating a halo. The scattering has two consequences: it removes photons from the direct line of sight (LOS) to the object, and it sends photons initially emitted out of the LOS towards the observer, creating the halo \citep[][]{predehl.2000fk}. With an extraction region of $50''$ radius, we can assume that these two effects compensate each other, so we do not include the loss due to dust scattering in our model. We note that we find a column density and a photon index consistent with past studies, a cross check on this assumption.

Figures \ref{fig:flares_spectrum} shows the spectra measured for the four flares with their best fit absorbed power-law model. The spectral binning is identical for each spectrum: spectra of FPM A and B are combined, and the spectral bins are grouped to reach at least 50 source $+$ background counts and 3 $\sigma$ signal-to-noise significance in each group. The high energy bound of the last group is set to maximize the significance of the group. Finally, the last group is merged with the second to last if it does not reach 2 $\sigma$ significance. The last bin significance is 2.51 (8.52--21.00 keV), 2.79 (9.88--26.00 keV), 2.88 (22.76--60.00 keV) and 3.77 (20.16--79.00 keV) for flares J20, J21\_1, J21\_2, and O17, respectively. 

We assume that the column density is constant from flare to flare, and we perform a joint fit of the four flares with the column density tied together. We find a column density of $1.66^{+0.70}_{-0.61} \times 10^{23}$ cm$^{-2}$, with a reduced $\chi^2$ of 0.90 (52.21/58). For comparison, column densities measured during past flares range from $1.25 \times 10^{23}$ cm$^{-2}$ to $1.85 \times 10^{23}$ cm$^{-2}$ \citep[converting published values to be consistent with Verner {\it et al.} cross sections and Wilms {\it et al.}  abundances,][]{goldwurm.2003kx, porquet.2003bh,belanger.2005kx,porquet.2008uq,nowak.2012fk}, being all consistent with each other within the uncertainties (which is consistent with our assumption of the column density being constant from flare to flare).  Table \ref{tab:flux} gives the best fit photon indices and associated fluxes and luminosities for each flare. The photon indices we find are consistent with those reported in previous work \citep{porquet.2003bh,belanger.2005kx,porquet.2008uq,degenaar.2013cr,nowak.2012fk}.

Flare O17 residuals show possible evidence for a cut off around 20 keV. For a direct comparison between the spectra of flares J21\_2 and O17, which have similar peak luminosity, we produced another O17 flare spectrum with an upper energy limit of 60 keV, and the highest energy point still falls below the power-law. However, fixing the column density to the value found for the four flares, an F-test  shows that a cut off is not required by the data (probability of 90.2\%).

We split flare J21\_2 into four phases to look for spectral evolution (see Figure \ref{fig:zoom_lc}). The spectral binning is made in the same way as described above, and the column density is fixed to $1.66 \times 10^{23}$ cm$^{-2}$. Figure \ref{fig:photindex}  shows the evolution of the spectral index during flare J21\_2 (right panel), and the best fit photon index of the power-law for all four flares (left panel). There is no spectral evolution during flare J21\_2, as $\Gamma$ changes by significantly less than its error across the flare, which confirms past results  \citep{porquet.2008uq, nowak.2012fk}. 

However, the 90\% confidence ranges of flares J21\_1 and O17 do not overlap. We investigate the significance of this difference by plotting the confidence contours of both photon indices (Figure \ref{fig:contourplot}). The calculation is done similarly to the spectral analysis presented above: the four flares are fitted simultaneously with tied column density. We find that the two photon indices are different at the 95\% level.
We know, thanks to {\em Chandra}, that we caught flare O17 right at its peak (J. Neilsen, private communication), but, as previously noted, there should not be any bias due to spectral evolution. Although this result is only marginally significant, we note that it is the first time such a difference has been reported between two individual flares. \citet{degenaar.2013cr} noticed a low-significance difference by comparing one bright flare to the average of five weak flares. Interestingly, the bright flare was softer ($L_{\rm 2-10\,keV, peak} = (3.1 \pm 0.5) \times 10^{35}$ erg s$^{-1}$, $\Gamma=3.0 \pm 0.8$) than the average of the five weaker ones, which is opposite to what we observe. This discrepancy can be interpreted as a piece of evidence that \SgrA X-ray flares can have a range of photon indices without being correlated to luminosity, keeping in mind that both the present result and that of \citet{degenaar.2013cr} have low significance. We note that magnetic reconnection is known to create MHD turbulence that produces stochastic acceleration, which can result in a range of particle distribution slopes \citep{zharkova.2011vn}. This could explain flares of various photon indices.

\begin{deluxetable}{l c c c c c c c c}
\tabletypesize{\scriptsize}
\tablecaption{Photon index, fluxes and luminosities for each flare in the 2--10 keV  and 3--79 keV energy ranges.  \label{tab:flux}}
\tablewidth{0pt}
\tablehead{
Flare ID & 
\colhead{$\Gamma$} &
\colhead{$F_{2-10}^{\rm {abs}}$  } & 
\colhead{$F_{2-10}^{\rm {unabs}}$  } & 
\colhead{$L_{2-10}^{\rm {unabs}}$  } & 
\colhead{$F_{3-79}^{\rm {abs}}$   } & 
\colhead{$F_{3-79}^{\rm {unabs}}$  } &
\colhead{$L_{3-79}^{\rm {unabs}}$  }  &
\colhead{Strength}
\\
  & &  \multicolumn{2} {c}{ $(10^{-11}$ erg\,cm$^{-2}$\,s$^{-1})$} & $(10^{35}$ erg\,s$^{-1})$ &  \multicolumn{2} {c}{ $(10^{-11}$ erg\,cm$^{-2}$\,s$^{-1})$} & $(10^{35}$ erg\,s$^{-1})$  & 
}
\startdata
\vspace{0.15cm}
J\_20 	       & $2.69^{+0.98}_{-0.84}$ & $ 0.38^{+ 0.14}_{- 0.14}$ & $ 0.85^{+ 0.62}_{- 0.39}$ & $ 0.65^{+ 0.48}_{- 0.30}$ & $ 0.70^{+ 0.56}_{- 0.30}$ & $ 0.95^{+ 0.50}_{- 0.34}$ & $ 0.73^{+ 0.38}_{- 0.26}$  & $18.1^{+13.3}_{-8.4}$  \\
\vspace{0.15cm}
J21\_1 	       & $2.84^{+0.64}_{-0.54}$ & $ 0.51^{+ 0.12}_{- 0.12}$ & $ 1.20^{+ 0.63}_{- 0.41}$ & $ 0.92^{+ 0.49}_{- 0.31}$ & $ 0.86^{+ 0.34}_{- 0.25}$ & $ 1.20^{+ 0.33}_{- 0.30}$ & $0.92^{+ 0.25}_{- 0.23}$ & $25.5^{+13.7}_{-8.9}$  \\
\vspace{0.15cm}
J21\_2	       & $2.23^{+0.24}_{-0.22}$ & $ 0.82^{+ 0.09}_{- 0.09}$ & $ 1.63^{+ 0.43}_{- 0.32}$ & $ 1.25^{+ 0.33}_{- 0.24}$ & $ 2.22^{+ 0.37}_{- 0.33}$ & $ 2.72^{+ 0.34}_{- 0.32}$ & $ 2.08^{+ 0.26}_{- 0.24}$ &  $34.7^{+9.9}_{-7.2}$  \\
\vspace{0.15cm}
O17 	                & $2.04^{+0.22}_{-0.20}$ & $1.33^{+0.14}_{-0.13}$ & $2.52^{+0.64}_{-0.48}$  & $1.93^{+0.49}_{-0.37}$ & $4.41^{+0.72}_{-0.65}$ & $5.18^{+ 0.65}_{- 0.61}$ & $ 3.97^{+ 0.50}_{- 0.47}$  & $53.6^{+14.8}_{-10.8}$  \\
\hline \\
\vspace{0.15cm}
J21\_2a  		& $1.90^{+0.48}_{-0.41}$ & $0.48^{+0.13}_{-0.13}$ & $0.88^{+0.34}_{-0.28}$  & $0.68^{+0.26}_{-0.21}$ & $1.88^{+0.84}_{-0.65}$ & $2.15^{+0.81}_{-0.63}$ & $1.65^{+0.62}_{-0.48}$ & $18.7^{+7.5}_{-6.1}$ \\
\vspace{0.15cm}
J21\_2b  		& $2.31^{+0.19}_{-0.18}$ & $1.02^{+0.11}_{-0.11}$ & $2.06^{+0.28}_{-0.27}$ & $1.58^{+0.22}_{-0.20}$ & $2.56^{+0.40}_{-0.36}$  & $3.18^{+0.39}_{-0.36}$ & $2.43^{+0.30}_{-0.27}$ & $43.8^{+7.6}_{-6.4}$ \\
\vspace{0.15cm}
J21\_2b peak 	& $2.44^{+0.33}_{-0.29}$ & $1.27^{+0.21}_{-0.20}$ & $2.66^{+0.60}_{-0.54}$  & $2.03^{+0.46}_{-0.41}$  & $2.85^{+0.72}_{-0.60}$ & $3.64^{+0.70}_{-0.62}$ & $2.79^{+0.53}_{-0.47}$ & $56.6^{+14.1}_{-12.0}$ \\
\enddata
\tablecomments{ The second column gives the best fit photon index $\Gamma$. The fluxes are determined using the \texttt{cflux} convolution model.
The column density of $1.66^{+0.70}_{-0.61} \times 10^{23}$ cm$^{-2}$ was determined by jointly fitting flares J20, J21\_1, J21\_2 and O17 with the column density tied together. The column density for the breakdown of flare J21\_2 (in the lower part of the table) was fixed to $1.66 \times 10^{23}$ cm$^{-2}$. Both absorbed (noted abs) and unabsorbed fluxes (noted unabs) are reported. The luminosity calculations assume a distance of 8 kpc and isotropic emission. The strength (rightmost column) is defined as the ratio of the 2--10 keV unabsorbed flux to the 2--10 keV unabsorbed quiescent flux, $0.47^{+0.05}_{-0.03} \times 10^{-12}$ erg cm$^{-2}$s$^{-1}$, as reported in \citet{nowak.2012fk}. All errors are at the 90\% confidence level.  }
\end{deluxetable}

\section{SEDs and one-zone static models}
\label{sec:sed}

An important step in understanding the production of the flares is to determine the underlying X-ray emission mechanism.  Synchrotron emission produces a spectrum following a power-law. In the SB case, where the X-ray flare is produced by the same electron population that emits the NIR flare, a spectral break occurs in the UV or optical due to fast electron cooling \citep{dodds-eden.2009dq}. Alternatively, we consider two IC scenarios, SSC and EC, both involving the electrons producing the NIR synchrotron emission flare (the third IC scenario mentioned in the introduction is not implemented here as it produces a similar X-ray spectrum to the other two). All three mechanisms could originate either in a jet or in an accretion flow. It is not our ambition to perform detailed modeling of the flaring emission. Instead, we want to identify the dominant mechanism by using single-component, one-zone (all parameters are constant across the emission region), and static models following \citet{dodds-eden.2009dq}. In previous work, \citet{eckart.2012bh} and \citet{yusef-zadeh.2009fk} used more sophisticated dynamical models to fit multi-wavelength light curves. Our simpler approach is appropriate given our limited bandpass (i.e. X-rays only). The constraints that we derive are reasonably robust, although some of the inferred parameters can change in more sophisticated flare scenarios \cite[e.g., the magnetic field strength,][]{eckart.2012bh}.

We calculate the X-ray spectra of the IC models according to the equations 18 to 21 of \citet{chiaberge.1999kx} \cite[see also][]{rybicki.1979fk}. The free parameters for the SSC scenario are the electron density $n_e$, the size of the emitting region $R$, the magnetic field strength $B$, and the temperature of the electrons $T_e$. Assuming a thermal distribution of electrons, the spectrum due to synchrotron emission is calculated, which is then up-scattered by the same population of electrons. Double up-scaterring is not accounted for in the model.

The EC scenario has additional free parameters, but we only vary those that affect the X-ray emission: $n_e$, $T_e$, $R$, and the size of the sub-mm emitting region, $R_{\rm sub-mm}$. The sub-mm spectrum is taken from the \citet{yuan.2003qf} model fit to the quiescent \SgrA spectrum. Again, only one up-scattering is accounted for, and due to the absence of NIR constraint, no synchrotron emission is calculated (its intensity would only depend on $B$).

The free parameters for the SB case are the index of the power-law distribution of the electron energy spectrum $p$, the magnetic field $B$, the total number of electrons $N_e$ (which strongly depends on the minimum Lorentz factor $\gamma_{\rm min}$), and the injection timescale $\tau_{\rm inj}$ (time for $N_e$ electrons to be accelerated). 
The cooling break creates the following electron distribution:
\[
{n(\gamma)} = \left\{
\begin{array} {l l}
{\gamma^{-p}}   & \gamma_{\rm min} \leq \gamma < \gamma_c \\
{\gamma^{-(p+1)}} & \gamma_{c} \leq \gamma < \gamma_{\rm max} \\
\end{array}
\right.
\]
where $\gamma_c$ is the Lorentz factor of the cooling break. An electron population with power-law index $p$ creates a power-law radiation spectrum with index $(p-1)/2$. $\gamma_c$ is given by the balance between injection of energetic electrons and synchrotron cooling. We estimate $\tau_{\rm inj}$ based on the variability timescale observed during flare J21\_2, 100~s. The cooling time, $\tau_{\rm cool}$, depends on the magnetic field strength and on the energy of the electrons (the shortest cooling time is given by $\gamma_{\rm max}$). $\gamma_c$ is determined by the condition \cite[][]{begelman.1984cr}:
\[
{\tau_{\rm cool} = \tau_{\rm inj}} \iff {3 \over 4} {{8 \pi m_e c} \over {\sigma_{\rm T} \, \gamma_c \, \beta^2_e \, B^2}} = 100 \;{\rm ,}
\]
where $m_e$ is the electron mass, $c$ is the speed of light, $\sigma_{\rm T}$ is the Thomson scattering cross-section for electrons, and $\beta_e$ is the ratio of the magnetic energy density to the thermal energy density. Solving this equation gives $\gamma_c= 8\times 10^{6} \times B^{-2}$  (assuming equipartition, $\beta_e = 1$). This induces a break in the synchrotron spectrum at the frequency 
\[
\nu_c \approx 0.45 \gamma_c^2 \, \nu_{\rm gyr} \, \sin(\alpha) \approx 6.3 \times 10^7  \gamma_c^2 \,\left({B \over {50\,{\rm G}}}\right) \approx 6.4 \times 10^{14} \left({{50\,{\rm G}} \over B}\right)^3 {\rm Hz}
\]
where $\nu_{\rm gyr}$ is the electron gyration frequency and $\alpha$ is the angle between the electron motion and the magnetic field direction (assumed to be $90^{\circ}$ here). For flare J21\_2, with an injection timescale of 100~s and an assumed value of 50 G for the magnetic field strength, the break is in the visible.

\begin{figure}[h]
\centering
\includegraphics[width=0.5\textwidth]{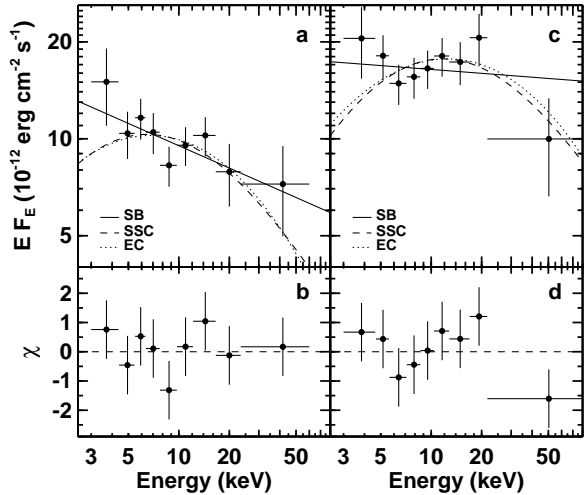}
\caption{Comparison of the {\em NuSTAR} SED for flare J21\_2 (a,b) and O17 (c,d) to three one-zone models: SB (solid line), SSC (dashed line), and EC (dotted line). See text for details. The vertical lines of the data points show the 1-$\sigma$ error bars. The lower panels show the residuals of the SB model.}
\label{fig:sed}
\end{figure}

The three aforementioned models were fit to the data using the chi-squared minimization method. Figure~\ref{fig:sed} compares the spectral energy distributions (SEDs) of flares J21\_2 and O17 to the three models. The SEDs were calculated by determining fluxes and errors for modules A and B separately and then averaging them. A coarse binning was chosen for the sake of clarity (see Appendix \ref{sec:appendixFT} for the significance of the bins).  We corrected the points for interstellar absorption as described in Appendix \ref{sec:appendixFT}. The best fit parameters for the three models are shown in Table~\ref{tab:SEDparam}.

\section{Implications of the SED measurements}
\label{sec:discussion}

\begin{deluxetable}{lcccccc}
\tabletypesize{\small}
\tablecaption{Best fit parameters for the EC, SSC and SB models of flares J21\_2 and O17.  \label{tab:SEDparam}}
\tablewidth{0pt}
\tablehead{Parameters & \colhead{EC$_{\rm J21\_2}$} & \colhead{EC$_{\rm O17}$} & \colhead{SSC$_{\rm J21\_2}$} & \colhead{SSC$_{\rm O17}$} & \colhead{SB$_{\rm J21\_2}$} & \colhead{SB$_{\rm O17}$}  }  
\startdata
$n_e$ (cm$^{-3}$)		& $4.2\times10^{7}$	 	& $3.7\times10^{8}$ 	& $3\times10^{10}$	  & $3.4\times10^{10}$ 	&  $> 2.9 - 1.9 \times 10^3$ 		& $> 5.7 - 5.6 \times 10^2$ \\
$R$ (cm)       			& $4.6\times10^{10}$	& $3.2\times10^{10}$ 	& $5.4\times10^{9}$   & $5.4\times10^{9}$ 	&  $ <  3 \times 10^{12}$	& $ <  3 \times 10^{12}$ \\
$R_{\rm sub-mm}$ (cm) 	& $1.1\times10^{11}$	& $6.4\times10^{10}$ 	&  ...				  & ... 				&   ...				& ... \\
$B$ (G)     			& ...					& ...					& 3200 			  & 2300  				&  $2.7 - 24$			& $6.3 - 56$ \\
$T_e$ (K)   			& $1.0\times10^{12}$	& $1.4\times10^{12}$	& $1.0\times10^{11}$ & $1.4\times10^{11}$ 	&   ...				& ... \\
$ \tau_{\rm inj} (s)$		& ...					& ... 					&  ...				  & ... 				&  100				& 100 \\
NIR flux (mJy)		     	& ...					& ...					&  5.5 			  & 5.1 				&  $1 - 27$			& $1 - 27$ \\
$\chi^2/{\rm dof}$  		& $7.3/7$				& $5.4/7$				& $7.2/7$			  & $4.8/7$ 			&  $4.1/7$				& $4.4/7$ \\
CDF					&  0.60  				&  0.39				& 0.59  			  & 0.31 				& 0.23  				& 0.27     \\
\enddata
\tablecomments{The NIR flux is the value for the $K_s$ band (2.2 $\mu$m); in the case of the SSC model, the NIR flux is the prediction from the model. In the case of the SB model, it is simply the range of past observations \citep{trap.2011fk}. The EC model does not allow a NIR flux prediction, as it only depends on the magnetic field strength given the number of electrons and their temperature. In the SB case, the number of electrons is calculated for $\gamma_{\rm min}=1000$ and $\gamma_{\rm max} =3 \times 10^5$. A range of electron densities and magnetic field strengths are quoted, which produce the same X-ray flux but cover the range of NIR flux measured in past flares. The size of the emitting region does not play a role.  The flux only depends on the total number of electrons. However, for consistency, we use the upper limit on the flaring region size that we derived for flare J21\_2 to estimate the lower limit on the electron density (see Section \ref{sec:energybalance}). The last row, CDF, gives the cumulative distribution function for the $\chi^2$ distribution, given the $\chi^2$ value and the degree of freedom (dof). }
\end{deluxetable}

Work to date combining NIR and X-ray data does not definitively rule out any of the above mechanisms. 
The narrow bandpasses  of {\em Chandra} and {\em XMM-Newton} and the strong absorption 
in the soft X-ray band have made it impossible to distinguish between a power-law spectrum,  
naturally explained by a synchrotron process, and a curved spectrum that would be characteristic of 
straightforward EC or SSC emission models. {\em NuSTAR}, with its larger bandpass, clearly shows that the spectrum extends to high-energy X-rays, but we are limited by statistics in the highest part of the energy range. The fit to flare J21\_2 is somewhat conclusive with a rejection probability for the SB model of 23\% versus 60\% and 59\% for the EC and SSC models, respectively. However, the fit to flare O17 is clearly inconclusive with no model standing out.

We nonetheless note that the best fit parameters for the EC and SSC models are unrealistic, while the SB model gives reasonable ranges of electron density and magnetic field strength. The EC process requires an electron density several orders of magnitude higher than expected for the accretion flow around \SgrA, and the high density required for the sub-mm photons leads to a very small sub-mm volume, over an order of magnitude smaller than the quiescent emission region as observed by VLBI \citep{fish.2011fk}. The SSC model needs even higher electron density along with extremely high magnetic field strengths. Thus, we reach identical conclusions to \citet{dodds-eden.2009dq}: we favor the SB model primarily based on physical arguments.

Within the synchrotron picture, {\em NuSTAR}'s high-energy detection implies either a higher magnetic field strength or a particle distribution with a higher maximum Lorentz factor ($E_{\rm max}\propto B\gamma_{\rm max}^{2}$) than previously assumed. With a magnetic field strength of 50~G, flare J21\_2 requires $\gamma_{\rm max}$$\sim$$3\times 10^{5}$, which implies a synchrotron cooling time as short as $\sim$1\,s for the most energetic X-ray emitting particles. Thus, continuous acceleration of high-energy particles is required to produce a $\sim$3,000\,s flare, even if the magnetic field were more than an order of magnitude lower than the value given above.

The rapid X-ray variability detected by {\em NuSTAR} also puts important
constraints on flare models. In the SB model with a cooling break between the NIR and X-ray,  
the NIR variability is primarily due to changes in the magnetic field strength, while 
any X-ray variability is driven by sporadic injection of high-energy particles 
into the emission region \citep{dodds-eden.2010fk}.  Thus, the amplitude changes (rise and decay) within 100~s measured during flare J21\_2 require that the particle injection rate can either drop or rise dramatically on a timescale of 100~s (see section \ref{sec:variability}).  Given that we observe these fast variations at the beginning and at the end of sub-flare J21\_2b (separated by 30 minutes), we assume that the emitting region keeps the same size throughout the flare. For this given flare, this assumption puts constraints on the adiabatically expanding blob model for the X-ray emission \citep{van-der-laan.1966kx,eckart.2006zr,eckart.2012bh,yusef-zadeh.2006fk,yusef-zadeh.2008uq,yusef-zadeh.2009fk}.

\section{Location of the emitting region: energy balance calculation}
\label{sec:energybalance}


Regardless of the emission mechanism and of the scenario considered (disk or jet), we estimate the amount of available energy for producing flares using the magnetic energy density ($B^{2}/8\pi$) and the volume of the flare's emission region ($V$). Considering theoretical models that predict magnetic field strengths  in the accretion flow and at the base of the jet, we compare the energy available at different distances from the SMBH to the fluence of the flare.

The natural speed associated with the magnetic energy density is the Alfven speed, but in the inner radii of the SMBH, this speed is likely to be close to the speed of light $c$. Since we seek an upper limit on the size of the emitting region, we use the speed of light as the propagation speed. With this assumption, the $\Delta \, t = 100$~s variability (rise and decay) observed during flare J21\_2 corresponds to an upper limit on the radial size of the X-ray emission region of $<~3.0~\times~10^{12}$~cm (2.5 Schwarzschild radii, $R_{\rm S}$), based on propagation at the speed of light from the center to the edge of a spherical region, disregarding the light travel time across the region. This corresponds to an upper limit on the emitting volume $V < (4/3)\pi c^{3} \Delta$$t^{3}$ and, in turn, on the magnetic energy, $U_{B} < B^{2}V/8\pi$, where $B$ is a function of the radial distance from the SMBH, $R$.  

The ion internal energy, which could amount to $\sim$10 times the magnetic energy \citep[e.g.][]{hirose.2004kx}, is not taken into account here as it is not clear how it would be converted into photons on such a short timescale. In any case, including the ion energy would require us to use the plasma sound speed as the relevant propagation speed (significantly smaller than $c$), which would compensate the effect of increased energy density by making the emitting volume much smaller and would lead to even tighter constraints on the flaring region location.

\begin{figure}[h]
\centering
\includegraphics[width=0.5\textwidth]{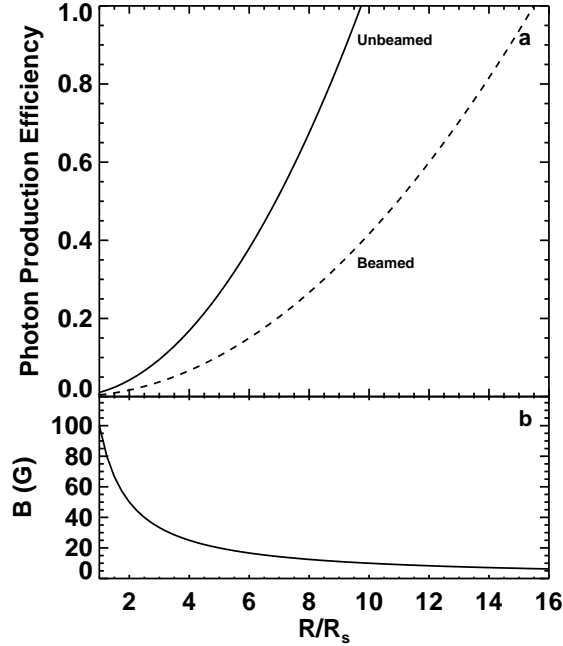}
\caption{Constraining the location of the flare based on the energy available in
the \SgrA disk or jet.  Panel {\em a} shows the required efficiency for converting
the available magnetic energy into the photons we see in flare J21\_2.  Panel {\em b}
shows our estimate (see text) for how the magnetic field strength ($B$) depends on 
the radial distance from the SMBH ($R$) in units of Schwarzschild radii
($R_{\rm S} = 2GM/c^{2}$, where $G$ is the gravitational constant, $M$ is the mass
of the SMBH, and $c$ is the speed of light).}
\label{fig:energybal}
\end{figure}

We estimate the strength and radial profile for $B$ based on theoretical considerations and previous accretion disk simulation results.  In the jet, a $R^{-1}$ dependence is predicted \citep{blandford_and_konigl_79}.  For a disk, the magnetic field strength is estimated from equipartition of magnetic and kinetic energy.  The theoretical dependence for a radiatively inefficient accretion flow (RIAF) is $B\propto R^{-1.05}$ and Bondi flows and Advection Dominated Accretion Flows give $B\propto R^{-1.25}$ \citep{yuan.2003qf,narayan.1998kx}, while simulations typically give between $R^{-0.7}$ and $R^{-1.2}$ \citep{dexter.2010uq}.  Thus, we conclude that the disk profile is very close to the jet profile, and we only show one curve, with $R^{-1}$, in Figure~\ref{fig:energybal}. Values in the inner part of the disk in the range of 10--100\,G have been quoted in previous work \citep{falcke.2000ys,yuan.2003qf,dodds-eden.2010fk}. Thus, we set the value to be $B_{R_{\rm S}} = 100$\,G at 1\,$R_{\rm S}$, and the exact profile we assume is shown in Figure~\ref{fig:energybal}b.  We view this as being a conservative assumption in the sense that a lower magnetic field would constrain the flare to occur even closer to the SMBH.

Figure~\ref{fig:energybal}a shows the main result, where we plot the photon production efficiency, $\eta$, which is the ratio of the fluence of the flare to the energy available at a given radial distance from the SMBH. As indicated by the fast rise and decay observed in Flare J21\_2, we assume that the emitting region keeps a constant size during the flare, and use the full duration of  the flare (1,898~s) and its average luminosity of $2.1\times 10^{35}$\,erg\,s$^{-1}$ for this calculation. Thus, based on energetics, there is only enough energy to power the flare if it originates within $< \,10 \, R_{\rm S} \, (B_{R_{\rm S}}/100$\,G)\,$\eta^{1/2}$ of the SMBH.

Here, we discuss how beaming could affect this conclusion. If the flare emission is beamed towards us, the rest frame variability timescale would be larger than the observed one, and the actual luminosity would be lower than we derived. Beaming becomes significant when the bulk motion of the emitting particles has a relativistic speed in a direction within a few tens of degrees from our line of sight. However, we argue here that this case is unlikely, and physically reasonable levels of beaming would have only a mild effect. 

The spin axis of the SMBH is largely unconstrained, but recent studies suggest that the spin axis points $>$35$^{\circ}$ 
away from our line of sight \citep{dexter.2010uq,broderick.2011zr,shcherbakov.2012fk}. In jet models, the X-ray emission comes predominately from the base of the jet, the so-called nozzle. The bulk speed in the nozzle is assumed to be the sound speed, $\sim 0.4c$ \citep{falcke.2000ys}. The Lorentz factor $\gamma$ of the bulk motion in the nozzle is thus $\sim1.1$, which yields a Doppler factor $D= 1/(\gamma \, (1 - \beta \cos \theta))$ of $\sim 1.4$ for the extreme case of $\theta=35^{\circ}$, very mildly affecting the timescales, and decreasing by a factor of 2 ($D^2$) the emitted luminosity (see consequences on Figure~\ref{fig:energybal}a).

Alternatively, one could think that a hot spot orbiting the SMBH could produce beamed emission if its orbital path were tangential to our line of sight. There is indeed evidence indicating that we may be seeing the disk nearly edge-on \citep{hamaus.2009kx}. However, typical orbital velocities are in the 0.3 -- 0.4~$c$ range, which does not produce strong beaming. Moreover, this configuration would produce a modulation at the orbital period (between $\sim 1860$~s and $\sim 360$~s at the innermost stable circular orbit for a SMBH that is non-rotating and maximally rotating, respectively), which has not previously been seen in X-ray observations.

\section{Summary}

Four flares were detected during this observation campaign, two of medium amplitude, J21\_2 and O17, and two weaker ones, J20 and J21\_1. For the first time, we clearly see that the power-law spectrum of \SgrA X-ray flares extends to high energy, with no evidence for a cut off; flare O17 is detected up to 79 keV and flare J21\_2 up to 60 keV. 

The SED fit slightly favors the SB model for flare J21\_2 but is inconclusive for flare O17 due to limited statistics at high energy. However, we confirm previous reports that physical arguments favor the SB model over the two simplified IC models we tested. This implies that efficient particle acceleration takes place continuously over the duration of the flares, as the synchrotron cooling time is of the order of 1~s. 

Flare J21\_2 was detected over a long enough time to search for spectral evolution, but we find none. We detect a variation of  photon index between two individual flares at the 95\% confidence level. Flare O17 ($\Gamma=2.04^{+0.22}_{-0.20}$), which is also stronger, is harder than flare J21\_1 ($\Gamma=2.84^{+0.64}_{-0.54}$). This is opposite to the finding of \citet{degenaar.2013cr} who reported on one bright flare that was softer than the average of five weak flares. Keeping in mind that the present result and that of \citet{degenaar.2013cr} have low significance, this indicates that the photon indices of flares can take a range of values, even for a given flux, which is suggestive of magnetic reconnection as acceleration mechanism.

Variability with a timescale of 100~s was observed during flare J21\_2, both in rise (factor of at least $3.8 \pm 1.1$) and decay (factor of $3.4 \pm 1.8$). This suggests that the flaring region keeps the same size throughout the flare, and it allows us to put an upper limit of 2.5 $R_{\rm S}$ on its radial size. Assuming that flares are powered by magnetic energy, this flaring region size places a constraint on the location of the region: energetics show that there is only enough energy to power flare J21\_2 within 10 $R_{\rm S}$ of the event horizon.

\acknowledgments

This work was supported under NASA Contract No. NNG08FD60C, and
made use of data from the {\it NuSTAR} mission, a project led by
the California Institute of Technology, managed by the Jet Propulsion
Laboratory, and funded by the National Aeronautics and Space
Administration. We thank the {\it NuSTAR} Operations, Software and
Calibration teams for support with the execution and analysis of
these observations.  This research has made use of the {\it NuSTAR}
Data Analysis Software (NuSTARDAS) jointly developed by the ASI
Science Data Center (ASDC, Italy) and the California Institute of
Technology (USA). The authors wish to thank S. Nayakshin, S. Markoff, A. Eckart, G. Trap, M. Wardle, and F. Yusef-Zadeh for useful discussions. 
We also thank the {\em Chandra} \SgrA XVP collaboration for information on absence of X-ray transients before and after the flares reported here.\\



{\it Facilities:} \facility{NuSTAR}.



\appendix

\section{Bayesian block search sensitivity}
\label{sec:appendix1}
\begin{figure}[h]
\centering
\includegraphics[width=0.5\textwidth]{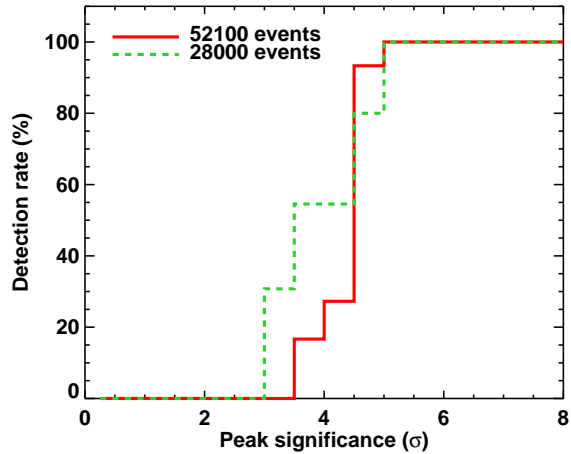}
\caption{Calibration of the Bayesian block peak search using synthetic event lists made of 52109 events (red solid line), and 28000 (green dash line). This plot shows the probability of detection of a rate increase lasting 1000s within a constant rate of 0.6 cts/s.}
\label{fig:BBsensitivity}
\end{figure}

In order to characterize the sensitivity of the peak search using the Bayesian block method (section \ref{sec:BB}), a set of synthetic event lists featuring a 1,000~s long rate increase were produced. The baseline rate was chosen to be 0.6 cts/s and the peak rate was gradually increased from 0.62 cts/s to 0.8 cts/s by steps of 0.02 cts/s. Fifteen event lists were generated per peak count rate value, making a total of 150 generated sets. The Bayesian block search was then applied to each event list, and the success of the search was evaluated by looking at the presence and location of the change points; Two change points, each within 1,000~s of the actual time of rise and decay is considered a successful search, even if more change points are detected. The actual peak significance is calculated using the knowledge of the actual rise and decay times, and 3,000~s before and 3,000~s after. It varies from set to set as a random generator is used to generate the intervals separating events, following a Poisson distribution. Finally, the trials are binned by actual peak significance, and for each bin, the ratio of the number of success to the number of trials is plotted. As we use a geometric prior for the Bayesian block search \citep{scargle.2013vn}, the sensitivity depends on the number of events. Figure \ref{fig:BBsensitivity} shows the results for 52,109 events and 28,000 events, which correspond to the July and October light curves, respectively.

This study shows that our settings ($fpr=0.01$, and geometric prior described in section \ref{sec:BB}) yield a detection chance $>$ 50\% for a peak significance between 4.5~$\sigma$ and 5~$\sigma$ and of 100\% for a peak significance $> 5~\sigma$. Performing these simulations with different peak duration leads to similar results, only the significance of the peak matters. So it appears that flare J20 detected during the July observation with a significance of 4.8 $\sigma$ is at the limit of detection. This corresponds to a strength of $\sim 14$ times the quiescent flux, for a flare of $\sim$ 920~s.

\section{Lightcurve of flare J21\_2}
\label{sec:J21_2}

\begin{figure}[h]
\centering
\includegraphics[width=0.5\textwidth]{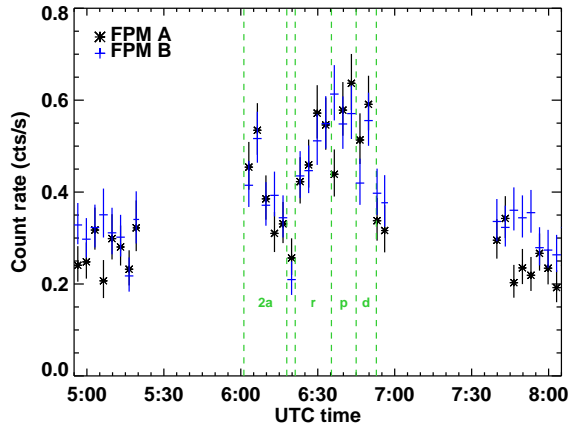}
\caption{Light curve of the J21\_2 flare (on 2012 July 21) with 200 s time resolution and 1-$\sigma$ upper and lower count rate limit. FPM A (black crosses) and FPM B (blue plus signs) are not combined in this plot. The data is extracted from a $50''$ radius extraction region centered on the radio position of \SgrA and is not background subtracted, as in Figures \ref{fig:full_lc} and \ref{fig:zoom_lc}.  The green vertical dash lines define the sub-flare (J21\_2a), and the rise (r), and peak (p), and decay phases of flare 21\_2b, as in Figures \ref{fig:zoom_lc} and \ref{fig:photindex}.}
\label{fig:J21_2}
\end{figure}

Figure \ref{fig:J21_2} shows again the 3--79 keV light curve of flare J21\_2 but without combining the two focal plane modules. The bin size is set to 200 s in order to yield comparable statistics with Figures \ref{fig:full_lc} and \ref{fig:zoom_lc}. One can see that the sharp rise and sharp decay are present in both modules (see section \ref{sec:variability}).

\section{Flares flux tables}
\label{sec:appendixFT}

The energy density for flares J21\_2 and O17 are given in Tables \ref{tab:Julyflux} and \ref{tab:Octflux}, respectively. We used the following method to calculate these SEDs. We start from the unfolded photon flux density (in photons cm$^{-2}$~s$^{-1}$~keV$^{-1}$), as returned by ISIS \citep{houck.2000kx}, and apply the absorption correction. Then, the energy density points are obtained by multiplying by $1.602 \times 10^{-9} \, E_{\rm geo}^2$, where $E_{\rm geo}$ is the geometric mean of each energy bin, defined as $E_{\rm geo} = \sqrt{E_1E_2}$. We use ISIS to unfold the spectrum {\it i.e.}, to account for the detector response matrix and the effective area of the optics.  For the absorption correction, we used the column density of $1.66^{+0.70}_{-0.61} \times 10^{23}$ cm$^{-2}$ that we determined with the joint fit of the four flares. The 90\% errors are divided by a factor of 1.7 to obtain the 1-$\sigma$ range $N_{\rm H}$ = 1.30--2.07$\times 10^{23}$\,cm$^{-2}$. For all three values ($1.30\times 10^{23}$, $1.66\times 10^{23}$, and $2.07\times 10^{23}$\,cm$^{-2}$), we calculated the ratio of unabsorbed to absorbed flux, and we multiplied the absorbed flux values by these ratios to make the absorption correction and to determine the errors on the absorption correction.

\begin{deluxetable}{lcccc}
\tabletypesize{\small}
\tablecaption{Spectral Energy Distribution for flare J21\_2.}
\tablewidth{0pt}
\tablehead{ \colhead{E (keV)} & \colhead{$\Delta E / 2$ (keV)} & \colhead{$E\,F_E$ (erg cm$^{-2}$ s$^{-1}$) } & \colhead{Error (erg cm$^{-2}$ s$^{-1}$)} & \colhead{Significance ($\sigma$) }}  
\startdata
   3.68     &  0.68    & 1.50314e-11  &  4.04299e-12 	& 7.78 \\
   4.92     &  0.52    & 1.04005e-11  &  1.73769e-12	& 7.36 \\
   5.90     &  0.42    & 1.16285e-11  &  1.68437e-12	& 7.48 \\
   7.02     &  0.66    & 1.04869e-11  &  1.52881e-12	& 7.09 \\
   8.72     &  1.00    & 8.27270e-12  &  1.17714e-12	& 7.29 \\
   10.94   &  1.18    & 9.54882e-12  &  1.31518e-12	& 7.29 \\
   14.36   &  2.20    & 1.02458e-11  &  1.42814e-12	& 7.45 \\
   19.98   &  3.38    & 7.90110e-12  &  1.74097e-12	& 4.55 \\
   41.70   &  18.30  & 7.23571e-12  &  2.25508e-12	& 3.02 \\
\enddata
\tablecomments{The first column gives the midpoint energy of the bin, the second, the half-width of the bin, the third, the specific flux multiplied by $E_{\rm geo}$, and the fourth, the 1-$\sigma$ error (including the propagation of the uncertainty on the column density).}
\label{tab:Julyflux}
\end{deluxetable}

\begin{deluxetable}{lcccc}
\tabletypesize{\small}
\tablecaption{Spectral Energy Distribution for flare O17.}
\tablewidth{0pt}
\tablehead{\colhead{E (keV)} & \colhead{$\Delta E / 2$ (keV)} & \colhead{$E\,F_E$ (erg cm$^{-2}$ s$^{-1}$) } & \colhead{Error (erg cm$^{-2}$ s$^{-1}$) } &\colhead{Significance ($\sigma$) } } 
\startdata
  3.82     &  0.82    &  2.05094e-11  &  5.12969e-12  & 7.70 \\
  5.14     &  0.46    &  1.81119e-11  &  2.86282e-12  & 7.55 \\
  6.38     &  0.74    &  1.48637e-11  &  2.11914e-12  & 7.44 \\
  7.88     &  0.72    &  1.55779e-11  &  2.26569e-12  & 7.40 \\
  9.50     &  0.86    &  1.65479e-11  &  2.25693e-12  & 7.55 \\
  11.60   &  1.20    &  1.80809e-11  &  2.47585e-12  & 7.38 \\ 
  14.88   &  2.04    &  1.73249e-11  &  2.64067e-12  & 6.59 \\
  19.28   &  2.32    &  2.06093e-11  &  3.82042e-12  & 5.42 \\
  50.32   &  28.68  &  9.99168e-12  &  3.36686e-12  & 2.98 \\
 \enddata
 \tablecomments{The columns description is identical to that in Table \ref{tab:Julyflux}}.
\label{tab:Octflux}
\end{deluxetable}

\clearpage




\bibliographystyle{jwapjbib}
\bibliography{refs}

\clearpage



\end{document}